\documentclass[fleqn,aps,pra,amsmath,amssymb,reprint,showpacs]{revtex4-1}

\usepackage{amsmath}
\usepackage{graphicx}
\usepackage{dcolumn}
\usepackage{bm}
\usepackage{epstopdf}

\begin{document}

\title{Ramsey fringes in a thermal beam of Yb atoms}

 \author{K. D. Rathod}
 \affiliation{Department of Physics, Indian Institute of
 Science, Bangalore 560\,012, India}
 \author{Vasant Natarajan}
 \affiliation{Department of Physics, Indian Institute of
 Science, Bangalore 560\,012, India}
 \email{vasant@physics.iisc.ernet.in}
 \homepage{www.physics.iisc.ernet.in/~vasant}

\begin{abstract}
We use the Ramsey separated oscillatory fields (SOF) technique in a $400^\circ$C thermal beam of Yb atoms to measure the Larmor precession frequency with high precision. For the experiment, we use the strongly-allowed ${^1S_0} \rightarrow {^1P_1}$ transition at $399$ nm, and choose the odd isotope $^{171}$Yb with nuclear spin $I=1/2$, so that the ground state has only two magnetic sublevels $m_F = \pm 1/2$. With a magnetic field of 22.2 G and a separation of about 400 mm between the oscillatory fields, the central Ramsey fringe is at 16.64 kHz and has a width of 350 Hz. The technique can be readily adapted to a cold atomic beam, and should be useful in experiments searching for a permanent electric dipole moment (EDM) in atoms.
\end{abstract}


\maketitle

\section{Introduction}
Ramsey's technique of separated oscillatory fields (SOF) \cite{RAM50} is widely used in atomic fountain clocks \cite{WYW05} and other high precision measurements \cite{WYC93,BOB97,CSP09}. This method is the workhorse for experiments searching for the existence of a permanent electric dipole moment (EDM) in a fundamental particle. Existence of an EDM is a signature of time-reversal symmetry violation in the fundamental laws of physics, and limits on EDM can serve to validate theories that go beyond the Standard Model (which is believed to be incomplete). EDM experiments have hence been done in the neutron \cite{BDG06}; paramagnetic atoms such as $^{133}$Cs \cite{MKL89} and $^{205}$Tl \cite{RCS02} (which imply the existence of an electron EDM); diamagnetic atoms such as $^{199}$Hg \cite{GSL09} (which are sensitive to the nuclear Schiff moment and time-reversal symmetry violating interactions); etc. Such experiments look for a shift in the Ramsey pattern in the presence of an electric field. Laser cooled Yb atoms, which have a spin-zero ground state and hence do not require a repumping laser, are prospective candidates for EDM experiments \cite{NAT05}. A {\em continuous} cold atomic beam has several advantages for these experiments; and recently we have demonstrated two ways of generating such beams \cite{RSN13,RPN13}. Implementation of Ramsey's SOF technique in such a cold beam of Yb atoms is an important requirement for reaching the desired precision in the EDM experiment.

In this work, we demonstrate the SOF technique in a {\em thermal} beam of Yb atoms, as a first step towards implementing it with cold atoms. The strongly-allowed ${^1S_0} \rightarrow {^1P_1}$ transition at $399$ nm, and the odd isotope $^{171}$Yb with nuclear spin $I=1/2$, are used in the experiment. $^{171}$Yb atoms emerging from a thermal source are first optically pumped to the $m_F = -1/2$ ground sublevel using a $\sigma^-$ pump beam. The atoms then pass through two rf interaction zones in a region that has a uniform quantization magnetic field that sets the Larmor precession frequency. The rf strength and interaction length of each zone is adjusted to give a $\pi/2$ pulse for atoms traveling with the mean velocity of 340 m/s. The two rf zones are separated by $\sim 400$ mm. The fraction of atoms left in $m_F = -1/2$ ground sublevel after the second interaction is measured using a $\sigma^+$ probe beam. The fluorescence signal from the probe as the rf frequency is scanned around the Larmor frequency shows a characteristic interference fringe pattern with a central fringe located at $16.64$~kHz having a linewidth of $350$ Hz. Both the lineshape and linewidth agree with the calculated probability of transition {\em averaged over the thermal velocity distribution}.

\section{Experimental details}
\begin{figure}
\centerline{\resizebox{1\columnwidth}{!}{\includegraphics{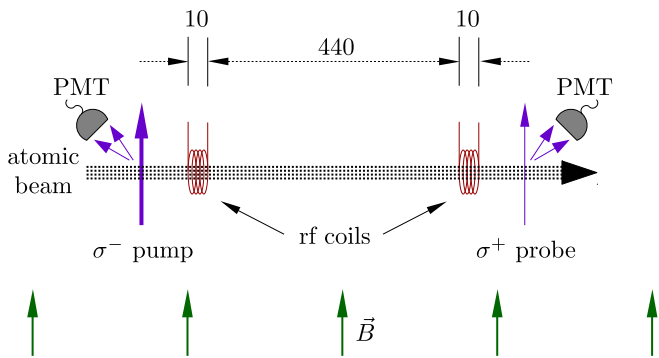}}}
\caption{(Color online) Schematic of the main parts of the experimental setup. A pair of rectangular coils (not shown), placed on either side of the atomic beam, produce the quantization field $B$.}
 \label{setup}
\end{figure}

The relevant parts of the set up are shown schematically in Fig.\ \ref{setup}. The experiments are done inside a ultra-high vacuum (UHV) chamber. The atomic beam is generated by resistively heating a quartz ampoule containing elemental Yb to a temperature of about $400^\circ$C. The source is not enriched and contains all isotopes in their natural abundances. The source part is maintained at a pressure of $10^{-7}$ torr by a $20$ l/s ion pump. It is connected to the main experimental chamber  consisting of a tube that is $500$ mm long with OD of $42$ mm, sandwiched between ports for optical access. This part is maintained at a pressure of $10^{-9}$ torr by a $75$~l/s ion pump. There is a differential pumping tube of 5~mm ID $\times$ $50$~mm length between the source and the experimental chamber. It serves the dual purpose of (i) allowing a pressure differential between the two regions, and (ii) collimating the atomic beam.

The atomic beam emerging from the differential pumping tube passes through a region with a constant magnetic field $B$ in the transverse direction. This field is produced by a pair of rectangular coils with dimensions $800$ mm $\times$ $130$ mm and separated by 160  mm, placed on the outside of the vacuum chamber. Each coil consists of $6$ turns of $8$ mm thick welding wire wound around four steel rods to provide shape and support. The coils carry a current $65$ A so as to produce a field of about $22$ G at the location of the atomic beam.

The required rf fields are produced using two coils wound around the outside of the 42 mm OD vacuum tube. Each coil is made using insulated copper wire of $0.6$ mm diameter, and consists of $90$ turns in $6$ layers. The physical width of the rf coils is $10$ mm and they are separated by $440$ mm, as shown in the schematic. The coils are driven with the output of a function generator amplified by a home-built audio amplifier.

The laser beams needed for the experiment are generated using a grating stabilized diode laser (Toptica DL100), operating with a Nichia diode at $399$ nm. The linewidth of the laser after feedback is about 1 MHz, and the total power available is 10 mW. The laser is locked to a particular magnetic transition (as described in the {\bf Results and Discussion} section) using current modulation at 20 kHz. The pump beam has $1/e^2$ diameter of $2.3$ mm and power $2.2$ mW, and is $\sigma^-$ polarized. It is used to spin polarize the unpolarized atoms coming out of the oven, by optically pumping into the $m_F = -1/2$ ground sublevel. The probe beam has $1/e^2$ diameter of $1$ mm and power of $50$ $\mu$W, and is $\sigma^+$ polarized. It is used to detect the population left in the $m_F = -1/2$ sublevel after the second rf interaction. All signals are fluorescence signals picked up with photomultiplier tubes (PMTs) from Hamamatsu (R928).

\section{Theoretical analysis}
The Zeeman shift (in Hz) of a sublevel with a given value of $m_F$ is
\begin{equation}
\delta\nu = g_F\mu_Bm_FB
 \label{shift}
\end{equation}
where $g_F$ is the $g$ factor of the level, $\mu_B = 1.4$ MHz/G is the Bohr magneton, and $B$ is the magnetic field in G. For the hyperfine level $F$, $g_F$ is given by \cite{MES99},
\begin{equation}
\begin{aligned}
g_F &= g_J\frac{F(F+1)+J(J+1)-I(I+1)}{2F(F+1)}  \\
 &-g_I\frac{\mu_N}{\mu_B}\frac{F(F+1)+I(I+1)-J(J+1)}{2F(F+1)}
 \label{gf}
\end{aligned}
\end{equation}
where $J$ is the total electronic angular momentum, $I$ is the nuclear spin, $g_J$ is the Land\'e $g$ factor, $g_I$ is the nuclear $g$ factor, and $\mu_N$ is the nuclear magneton.

\begin{figure}
\centerline{\resizebox{.85\columnwidth}{!}{\includegraphics{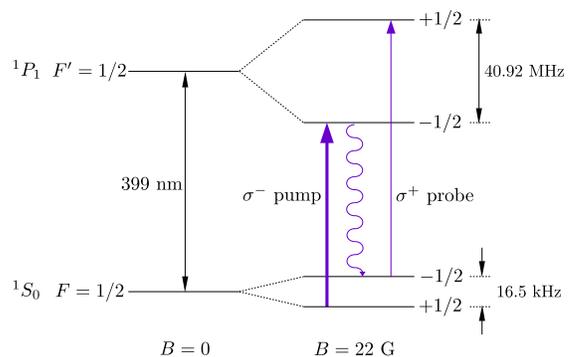}}}
\caption{(Color online) Splitting of the ground $^1S_0$ level and the excited $^1P_1$ level in $^{171}$Yb in the presence of a magnetic field $B$. Note that the splitting is opposite for the ground level because $g_F$ is negative. The transitions coupled by the $\sigma^-$ pump beam and the $\sigma^+$ probe beam are shown. The wavy line represents spontaneous decay.}
 \label{levels}
\end{figure}

For the $^1S_0$ ground state in $^{171}$Yb, $J$ is $0$, so the first term in Eq.\ \eqref{gf} is 0 and only the second term contributes. The nuclear magnetic moment $\mu_n$ for $^{171}$Yb is $+0.4926 \, \mu_N$ \cite{GJW64}. Using $\mu_n = g_I\mu_NI$, the $g$ factor for the $F=1/2$ level is
\[
g_F = -5.37 \times 10^{-4}
\]
From Eq.\ \eqref{shift}, this gives a shift of $-375$ Hz/G for the $m_F = + 1/2$ sublevel. \\
For the excited $^1P_1$ state, $J$ is 1, so both terms in Eq.\ \eqref{gf} are potentially important. But we can ignore the second term since $\mu_N/\mu_B = 1/1836$. Thus the $g$ factor for the $F'=1/2$ level is
\[
g_{F'} = +1.33
\]
This gives a shift of $+0.93$ MHz/G for the $m_{F'} = +1/2$ sublevel, which is opposite in sign to the ground one because of the opposite signs of the $g$ factors. The splitting of these two levels in a magnetic field of $22$ G is shown in Fig.\ \ref{levels}. The ground level splitting gives the Larmor precession frequency, so it is $16.5$ kHz for this field.

Once all the atoms are pumped into the $m_F = -1/2$ sublevel, there is a probability $p$ that they have made a transition to the $m_F = +1/2$ sublevel after interacting with the two rf fields. If the two rf zones have a length $l$ and are separated by a distance $L$; a particular atom has a velocity $v$; and the rf field has a strength $\Omega_R$ (its Rabi frequency) and its oscillation frequency is detuned from resonance by $\Delta$; the transition probability is given by \cite{RAM50},
\begin{equation}
\begin{aligned}
p &= 4 \, \dfrac{\Omega^2_R}{\Omega'^2_R} \,
\sin^2 \left( \dfrac{\Omega'_R l}{2v} \right)\times \\
& \, \ \ \left[ \cos\dfrac{\Delta L}{2v}
\cos\dfrac{\Omega'_R l}{2v} -\dfrac{\Delta}{\Omega'_R}\sin\dfrac{\Delta L}{2v}\sin\dfrac{\Omega'_Rl}{2v}\right]^2
 \label{prob}
\end{aligned}
\end{equation}
where  $\Omega'_R = \sqrt{\Omega_R^2 + \Delta^2}$ is the effective Rabi frequency.
Since the atoms emanate from a thermal source, the beam consists of atoms with the following longitudinal velocity distribution \cite{RAM56}
\begin{equation}
f(v) = 2\left(\frac{m}{2k_BT}\right)^2v^3\exp\left(-\frac{mv^2}{2k_BT}\right)
\label{dist}
\end{equation}
so that $f(v)dv$ is the number of atoms with velocity between $v$ and $v+dv$. The total probability $P$ in an atomic beam is thus the probability for one velocity given in Eq.\ \eqref{prob} averaged over the velocity distribution given by Eq.\ \eqref{dist}. Hence
\begin{equation}
P = \int_{0}^{\infty}p \, f(v)dv
\end{equation}
Since in the experiment we detect the population that has stayed in the $m_F = -1/2$ sublevel, the fluorescence signal is proportional to $(1-P)$. The calculated probability is compared to the observed signal in the next section.

\section{Results and discussion}
\begin{figure}
\centerline{\resizebox{0.95\columnwidth}{!}{\includegraphics{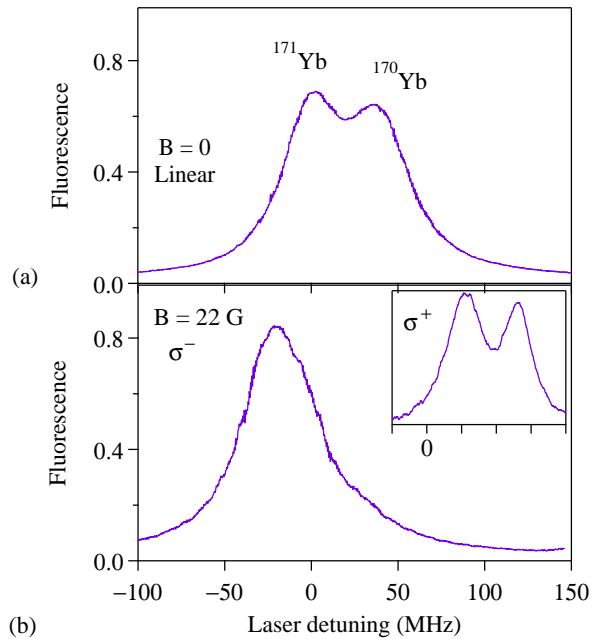}}}
\caption{(Color online) Scheme for locking the laser (a) The peaks corresponding to the $^{171}$Yb ($F=1/2 \rightarrow F'=1/2$) transition and the $^{170}$Yb ($F=0 \rightarrow F'=1$) transition are partially resolved in zero field, preventing proper locking of the laser. (b) There is only a single peak in the presence of a magnetic field and when the laser beam has $\sigma^-$ polarization; the peak is actually an unresolved peak consisting of both the $^{171}$Yb ($m_F=+1/2 \rightarrow m_{F'}=-1/2$) transition and the $^{170}$Yb ($m_F=0 \rightarrow m_{F'}=-1$) transition, which is unimportant as explained in the text. The inset shows that choosing $\sigma^+$ polarization also leads to two partially resolved peaks, corresponding to the $^{171}$Yb ($m_F=-1/2 \rightarrow m_{F'}=+1/2$) transition and the $^{170}$Yb ($m_F=0 \rightarrow m_{F'}=+1$) transition. All peaks are shifted from 0 because of the magnetic field.}
\label{lock}
\end{figure}

For the optical pumping to work effectively, the pump laser needs to be locked as close to the $m_F=+1/2 \rightarrow m_{F'}=-1/2$ transition in $^{171}$Yb (see Fig.\ \ref{levels}) as possible. But this is not as straightforward as it seems because of the  presence of the overlapping peak from the even isotope $^{170}$Yb. This will affect the lock point significantly because the two peaks are partially resolved, as seen from the no field spectrum shown in Fig.\ \ref{lock}(a). This problem can be solved by using a magnetic field to shift the sublevels, and using circularly polarized light so that only a subset of transitions are allowed. But with $\sigma^+$ polarized light, this still leads to an issue because the final spectrum has two peaks, as shown in the inset of Fig.\ \ref{lock}(b). On the other hand, with $\sigma^-$ polarized light, there is only one peak, as seen in Fig.\ \ref{lock}(b). This peak is actually an unresolved peak containing both the $+1/2 \rightarrow -1/2$ transition in $^{171}$Yb and the $0 \rightarrow -1$ transition in $^{170}$Yb, but the presence of this additional transition is unimportant because optical pumping only requires the pump laser to be on the correct transition. This is one of the main reasons for choosing $\sigma^-$ polarization for the pump beam.

\begin{figure}
\centerline{\resizebox{1\columnwidth}{!}{\includegraphics{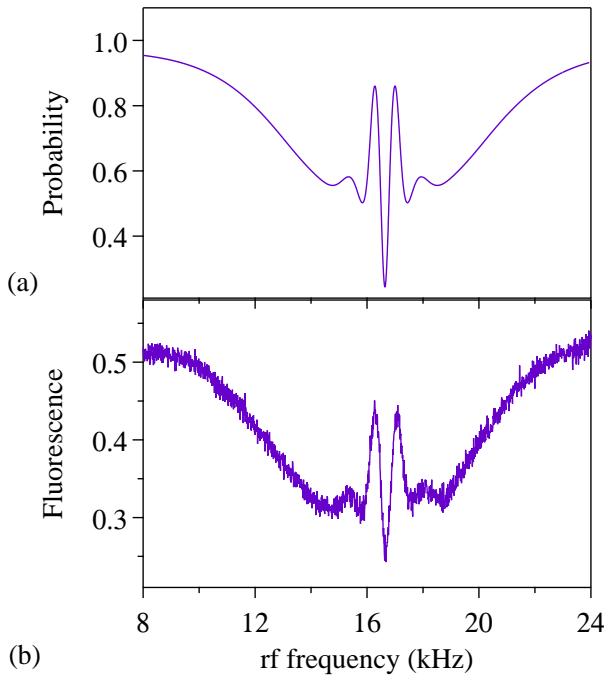}}}
\caption{(Color online) Theoretical and experimental Ramsey fringe spectra as a function of rf frequency. (a) Calculated probability of atoms being in the $m_F = -1/2$ sublevel, averaged over the thermal velocity distribution. (b) Measured fluorescence signal from the $\sigma^+$ probe beam, corresponding to population in the $m_F = -1/2$ sublevel. The central fringe in both cases has a width of 350 Hz.}
 \label{fringes}
\end{figure}

The experimentally observed Ramsey pattern is compared to the theoretical calculation in Fig.\ \ref{fringes}. The theoretical calculation is done according to the procedure discussed in the previous section. The coil length $l$ and the free evolution distance $L$ are adjusted to get a good fit. The geometrical dimensions are $l=10$ mm and $L=440$ mm (see Fig.\ \ref{setup}), whereas the best fit is obtained with values of 40 mm and 400 mm. This increase in $l$ (and corresponding decrease in $L$) is reasonable because of the fringing fields from the coils extending beyond their physical edge, and the fact that their mean diameter is much larger than the size of atomic beam. Changing the oven temperature [which determines the velocity distribution in Eq.\ \eqref{dist}] has negligible effect on the calculation, and it is left at $400^\circ$C.

As seen in the figure, the lineshapes for theory and experiment match quite well. The central fringe is located at $16.64$ kHz corresponding to a magnetic field of 22.2 G, which is in good agreement with the value calculated from the coil dimensions. The linewidth for the central fringe in both theory and experiment is about $350$ Hz.

\section{Conclusion}
In summary, we have demonstrated Ramsey's SOF technique in a thermal beam of Yb atoms emanating from an oven at $400^\circ$C. The experiment was done on the $^1{S_0} \rightarrow {^1P}_1$ transition at $399$ nm, and using the isotope $^{171}$Yb which has $I=1/2$ and hence just two magnetic sublevels in the ground state. Using a quantization field of 22.2 G, we obtain a linewidth of $350$ Hz at 16.64 kHz for the central fringe in the Ramsey pattern. The line shape and width agree with the theoretically calculated transition probability, after averaging over the thermal velocity distribution.

\begin{figure}
\centerline{\resizebox{1\columnwidth}{!}{\includegraphics{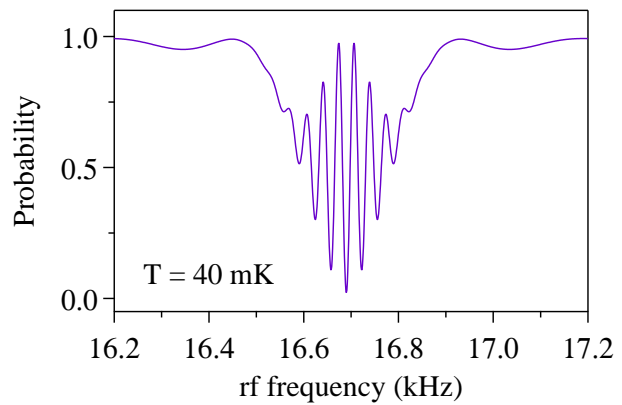}}}
\caption{(Color online) Calculated Ramsey pattern for a cold atomic beam with a longitudinal temperature of 40 mK. All parameters are the same as that used for Fig.\ \ref{fringes}, except for the rf strength as discussed in the text. The scan range is only 1 kHz, and the central fringe has a width of 15 Hz. }
 \label{cold}
\end{figure}

We now plan to implement this in a cold atomic beam, as demonstrated by us in earlier work \cite{RPN13}. In that case, the beam consisted of atoms traveling with a mean velocity of 15 m/s at a longitudinal temperature of 40 mK. To see the improvement that is likely, we have repeated the calculation for this temperature. The values of $B$, $l$, and $L$ are left unchanged, only the rf strength is adjusted to provide a $\pi/2$ pulse for atoms traveling at 15 m/s. The results are shown in Fig.\ \ref{cold}. Note that the scan range is only 1 kHz, which is 16 times smaller than for the previous one. The number of fringes has increased, and there is a dramatic decrease in linewidth of the central fringe from 350 Hz to 15 Hz. This illustrates the advantage of using a cold beam for EDM experiments.

\begin{acknowledgments}
This work was supported by the Department of Science and Technology, India, through the Swarnajayanti fellowship. The authors thank Sambit Pal for help with an initial version of this experiment, and for designing the audio amplifier. K.D.R. acknowledges financial support from the Council of Scientific and Industrial Research, India.
\end{acknowledgments}

%

%

\end{document}